\documentclass[12pt]{article}
\usepackage{tabularx}
\usepackage{array}
\usepackage{graphics}
\usepackage{graphicx}
\usepackage{epsfig}
\usepackage{amsmath}
\usepackage{citesort}
\usepackage{a4wide}

\makeatletter

\newcommand{\nn}{\nonumber}
\newcommand{\bea}{\begin{eqnarray}}
\newcommand{\eea}{\end{eqnarray}}
\newcommand{\be}{\begin{equation}}
\newcommand{\ee}{\end{equation}}
\newcommand{\gev}{\,{\rm GeV}}

\newcommand{\mev}{\,{\rm MeV}}

\newcommand{\pdir}{p\kern -5.2pt\raise 0.2ex\hbox {/}}
\newcommand{\vdir}{v\kern -5.75pt\raise 0.15ex\hbox {/}}
\newcommand{\kdir}{k\kern -5.75pt\raise 0.15ex\hbox {/}}
\newcommand{\epsdir}{\epsilon\kern -5.0pt\raise 0.15ex\hbox {/}}
\newcommand{\bvdir}{\bar{v}\kern -5.75pt\raise 0.15ex\hbox {/}}
\newcommand{\Ddir}{D\kern -7.75pt\raise 0.20ex\hbox {/}}
\newcommand{\ldir}{l\kern -5.0pt\raise 0.2ex\hbox{/}}
\newcommand{\varepsdir}{\varepsilon\kern -5.5pt\raise 0.15ex\hbox{/}}

\newcommand{\g}{\gamma}

\newcommand{\xii}{3549(13)(19)(92)}
\newcommand{\xiis}{3641(18)(8)(95)}
\newcommand{\om}{3663(11)(17)(95)}
\newcommand{\oms}{3734(14)(8)(97)}
\newcommand{\dsigcd}{18(51)}
\newcommand{\dxicd}{30(33)}
\newcommand{\domcd}{43(17)}
\newcommand{\dxiccd}{89(15)}
\newcommand{\domccd}{69(10)}
\newcommand{\dsigcr}{49(39)(12)(1)}
\newcommand{\dxicr}{47(27)(4)(1)}
\newcommand{\domcr}{44(16)(15)(1)}
\newcommand{\dxiccr}{87(13)(13)(2)}
\newcommand{\domccr}{67(9)(13)(2)}

\long\def\@makecaption#1#2{%
  \vskip\abovecaptionskip
  \sbox\@tempboxa{\small{\bfseries #1.} \  #2}%
  \ifdim \wd\@tempboxa >\hsize
    \small{\bfseries #1.} \  #2\par
  \else
    \global \@minipagefalse
    \hb@xt@\hsize{\hfil\box\@tempboxa\hfil}%
  \fi
  \vskip\belowcaptionskip}

\def\lsim{\mathrel{\lower .3ex\hbox{$\buildrel\textstyle<\over\sim$}}}
\makeatother

\begin{document}
\thispagestyle{empty} 
\begin{flushright}
\begin{tabular}{l}
{\tt SHEP-0319}\\
\end{tabular}
\end{flushright}

\begin{center}
\vskip 2.2cm\par
{\par\centering \LARGE \bf Spectroscopy of Doubly-Charmed\\
Baryons in Lattice QCD }\\   
\vskip 1.30cm\par
{\par\centering \large
UKQCD Collaboration\\[0.8ex]  
J.M.~Flynn, F.~Mescia and A.S.B.~Tariq}
{\par\centering \vskip 0.5 cm\par}
{\sl 
Department of Physics and Astronomy, University of Southampton, \\
Southampton, SO17 1BJ, United Kingdom.}

{\vskip 0.75cm\par}
\end{center}

\begin{abstract}
We present results for masses of spin-${1/2}$ and spin-${3/2}$
double-charm baryons in quenched lattice QCD, from an exploratory
study using a non-perturbatively improved clover action at
$\beta=6.2$. We have studied local operators and we observe, after
appropriate projections, a good signal for the ground states. We also
present results for single-charmed baryons and spin-splittings for
both double- and single-charmed states.
\end{abstract}
\noindent

\renewcommand{\thefootnote}{\arabic{footnote}}

\newpage
\setcounter{footnote}{0}

\section{Introduction}
\setcounter{equation}{0}

Recently SELEX, the charm hadroproduction experiment at Fermilab, has
reported a narrow state at $3519\pm 1\pm 5\mev$. This state decays
into $\Lambda_c^+K^-\pi^+$, consistent with the weak decay of the
doubly charmed baryon $\Xi_{cc}^+$~\cite{Mattson02}. This is the first
observation of a baryon containing two charm quarks. BaBar and Belle
at the SLAC and KEK $b$-factories may also provide further evidence of
doubly charmed baryons.

Double charmed baryons combine the opposites of the slow relative
motion of two heavy quarks with the fast motion of a light quark. They
provide scope for testing ideas developed for single charm physics,
such as the predicted hierarchies in lifetimes and semi-leptonic
branching ratios and give us more room to explore predictions of
exotic tetra- and penta-quark states (see the review by
Richard~\cite{Ric02} for further details). Unresolved issues regarding
the recent SELEX observation itself include the fact that the observed
lifetime of less than $30\,\mathrm{fs}$ is much less than predicted by
quark models~\cite{KL02} and that the observed isospin splitting
$m(ccd)-m(ccu)$ is rather large, about 60 times that for nucleons. The
latter is, perhaps, the cause for publication of only one of the three
states (the others being around $3460$ and $3780\mev$) observed in the
first instance (see e.g.~\cite{Moi02}). Therefore it is important to
study these baryons further.

The first prediction for the masses of these double-charmed baryons
comes from the early work of De Rujula {\em et al.}~\cite{DeR75}, with
later calculations from quark models and QCD sum rules
\cite{FR89,SW90,Bag94,Ron95,Ger99,Ton00,KK01,Ebe02,NT02,Nar02}.

Lattice QCD provides a method of calculating the masses of these
baryons from first principles in a model-independent and
non-perturbative manner. It is interesting to compare the results from
different lattice calculational techniques. Previous lattice
calculations have used the D234 action~\cite{Lew01} and
NRQCD~\cite{Mat02}. NRQCD is less suitable for charm quarks than for
beauty quarks, and furthermore charm quark masses are very accessible
to lattice simulation without using an effective theory. In this
calculation we use a non-perturbatively $\mathcal{O}(a)$-improved
clover action~\cite{Lus96}. Thus our results have discretisation
errors $\mathcal{O}(m_q^2 a^2)$, with $m_c^2 a^2$ about $25\%$ in our
simulation. Interestingly, although our double-charm baryons have
masses larger than one in lattice units, the physical mass looks
consistent for the one state for which an experimental number is
available.

We also study spin-splittings for charmed baryons and mesons, where
the leading charm quark mass dependence cancels. Recent calculations
using the $\mathcal{O}(a)$ non-perturbatively improved clover action
find vector-pseudoscalar meson splittings in better
agreement~\cite{Bec98,Bow00} with experiment than earlier calculations
using less-improved clover actions~\cite{All92}. We confirm this here.
For the single-charmed baryons, calculations with a tree-level clover
action had difficulty reproducing the experimental
splittings~\cite{Bow96}, while simulations using the D234~\cite{Lew01}
or NRQCD~\cite{Mat02} actions were compatible with experiment. We too
find compatibility. For the doubly-charmed baryons, experimental data
are not yet available. However, our results are compatible with those
from the D234 and NRQCD actions. Since the hyperfine splitting is
sensitive to the chromomagnetic moment term in the improved clover
fermion action, we believe this shows the importance of using the
non-perturbative value for its coefficient ($c_\mathrm{SW}$). A
similar observation was made concerning the coupling with the
chromomagnetic field in the NRQCD action~\cite{Mat02} ($c_4$ in
eq.~(A5) in~\cite{Mat02}).

Our final results for the double-charm masses and splittings are
\be
\begin{array}{rclrcl}
\Xi_{cc}&=&\xii\,\mev&\quad\Omega_{cc}&=&\om\,\mev \\
\Xi^*_{cc}&=&\xiis\,\mev&\quad\Omega^*_{cc}&=&\oms\,\mev \\
\Xi_{cc}^* - \Xi_{cc} &=& \dxiccr\,\mev&\quad
\Omega_{cc}^*-\Omega_{cc} &=& \domccr\,\mev
\end{array}
\ee
The splittings are determined by fitting ratios of
correlators, which gives smaller errors compared to taking a
difference of separately-fitted masses. Results for charmed meson
splittings and single-charm baryon masses and splittings are in the
body of the paper.

\begin{table}
\hbox to\textwidth{\hss
\begin{tabular}{|ccc|}
\hline
\vrule height2.5ex width0pt
Baryon & Quark content   &     Mass [MeV] \\[0.5ex]
\hline
\multicolumn{3}{|c|}{$s_{cc}=1,\,J^P={1/2}^{+}$ }\\\cline{2-2}
$\Xi_{cc}$ & $ccu$, $ccd$ &  3519(5)\\
$\Omega_{cc}$     &  $ccs$  & \\
\hline
\multicolumn{3}{|c|}{$s_{cc}=1,\,J^P={3/2}^{+}$ }\\\cline{2-2}
$\Xi^{*}_{cc}$    &  $ccu$, $ccd$  & \\
$\Omega^{*}_{cc}$ &  $ccs$  & \\
\hline
\multicolumn{3}{|c|}{$s_{ll}=0,\,J^P={1/2}^{+}$ }\\\cline{2-2}
$\Lambda_{c}$ & $cud$  &  $2285(1)$\\
$\Xi_{c}$     & $cus$, $cds$  &  $2469(1)$\\
\hline
\multicolumn{3}{|c|}{$s_{ll}=1,\,J^P={1/2}^{+}$ }\\\cline{2-2}
$\Sigma_{c}$  	& $cuu$, $cud$, $cdd$   & $2452(1)$\\
$\Xi'_{c}$ & $cus$, $cds$ & $2575(3)$\\
$\Omega_{c}$    &  $css$  & $2698(3)$\\
\hline
\multicolumn{3}{|c|}{$s_{ll}=1,\,J^P={3/2}^{+}$ }\\\cline{2-2}
$\Sigma^{*}_{c}$ &  $cuu$, $cud$, $cdd$  & $2518(2)$ \\
$\Xi^{*}_{c}$    &  $cus$, $cds$  & $2646(2)$\\
$\Omega^{*}_{c}$ &  $css$  & \\
\hline
\end{tabular}
\hss}
\caption{Summary of charmed baryons. Valence quark content and
spin-parity are shown.  The quantities $s_{cc}$ and $s_{ll}$ are the
total spin of the charm and light quark pair respectively.  The
experimental values are from ref.~\cite{pdg}, averaged over isospin
multiplets. The $\Xi_{cc}$ mass is from the recent observation of the
$\Xi_{cc}^+(ccd)$~\cite{Mattson02}.}
\label{tab:Expc}
\end{table}

In this paper the theoretical input and computational details of
the simulation are given in Sections~\ref{sec:states}
and~\ref{sec:details}, whereas the results are presented and analysed
in Section~\ref{sec:results}.  There is a brief conclusion in
Section~\ref{sec:conclusion}.

\section{Baryon states and interpolating operators }
\label{sec:states}

The double and single charmed baryons expected in QCD are summarised
in tab.~\ref{tab:Expc}.

On the lattice, the masses of these hadrons can be calculated in the
usual way from the large time behaviour of two point correlation
functions \be C(t)=\sum_{\bf x}\langle 0\vert J({\bf
x},t)\bar{J}(0)\vert 0\rangle
\label{eq:gen}
\ee
where the $J$'s are interpolating operators with quantum numbers
to create or annihilate the state of interest.  The choice of
operators is not unique.

For the spin-$1/2$ doubly-heavy baryon states, a simple operator is
\begin{equation}
J_\gamma= 
\epsilon_{abc}\,h_{\gamma}^{a}\,
\left({h^{b}}^T\,\gamma_5 {\cal C} \,l^{c}\right),\qquad s_{hh}=1
\label{eq:hhl5}
\end{equation}
where $a,\,b,\,c$ are colour indices, ${\cal C}$ is the charge
conjugation matrix and the $h$ and $l$ fields stand for generic
heavy and light quarks.

In $S$-wave baryons with two identical quarks (heavy quarks in our
case), the two quarks cannot couple to spin zero and the only
possibility is $s_{hh}=1$ (symmetric in both spin and flavour). The
component $s_{hh}=0$ as well as the operator
$-\epsilon_{abc}\,l_{\gamma}^{a}\,({h^{b}}^T\,\gamma_5{\cal C}
\,h^{c})$ vanish. The coupling of the light-quark spin to $s_{hh}=1$,
however, can also generate the spin $3/2$ states, $\Xi^*_{hh}$ and
$\Omega^*_{hh}$ in tab.~\ref{tab:Expc}.

An interpolating operator for the spin-$3/2$ states can be obtained by
replacing $\g_5$ with $\g^\mu$ in eq.~(\ref{eq:hhl5}).
\begin{equation}
J^{\mu}_\gamma = \epsilon_{abc}\,h_{\gamma}^{a}\,
\left({h^{b}}^T\,\gamma^\mu{\cal C} \,l^{c}\right),\quad s_{hh}=1.
\end{equation}
This operator also couples to spin-$1/2$ and projections are needed to
obtain the desired state. The spin-$1/2$ masses from $J^\mu_\g$ and
$J_\g$ are equal since there is only one spin $1/2$ baryon in the
situation where two quarks are identical. We have directly verified
this property in our simulation.

Another operator, used for spin-$3/2$ double heavy
baryons~\cite{Mat02,Ali00} is
\be \widetilde
J^{\mu}_{\gamma}=\epsilon_{abc}\,l_{\gamma}^{c}\,({h^{b}}^T\,\gamma^\mu{\cal
C} \,h^{a}).  
\ee 
We have also tried this operator and we see no reason to prefer one
over the other. Indeed, both give a good overlap for the ground state
and the masses extracted turn out to be equal as expected.

For the operators $J_\gamma$ and $J^\mu_\gamma$ (or $\widetilde
J^{\mu}_{\gamma}$) the $2$-point functions in
eq.~(\ref{eq:gen}) have the following large-time behaviour 
\bea
\label{eq:c2tlarge}
C(t)_{\gamma\bar{\gamma}}& = &\sum_{\bf x}
 \langle 0\vert J_\gamma({\bf x},t)
 \bar{J}_{\bar{\gamma}}(0)\vert 0\rangle \\
&\stackrel{t\gg0}{\longrightarrow}&
 Z_{1/2}\,
\left(P_+\right)_{\gamma\bar\gamma}\:e^{-m_{1/2}t}
+\, Z_{1/2}^P\,\left(P_-\right)_{\gamma\bar\gamma}\:
e^{-m_{1/2}^Pt}.\nonumber \\\nn\\
C_2(t)^{ij}_{\gamma\bar{\gamma}}&=&\sum_{\bf x}\langle 0\vert J^i_\gamma({\bf x},t)\bar{J}^j_{\bar{\gamma}}(0)\vert 0\rangle\\
& \stackrel{t\gg0}{\longrightarrow}&
\phantom{+}Z_{3/2}\, \left(P_+ \,P_{3/2}^{ij}\right)_{\gamma\bar{\gamma}}\:e^{-m_{3/2} t}+
Z_{1/2}\,\left( P_+\, P_{1/2}^{ij}\right)_{\gamma\bar{\gamma}}\:e^{-m_{1/2} t}\nn\\
&&
+Z_{3/2}^P\,\left(P_- \,P_{3/2}^{ij}\right)_{\gamma\bar{\gamma}}\:e^{-m^P_{3/2} t} +
Z_{1/2}^P\,\left( P_-\, P_{1/2}^{ij}\right)_{\gamma\bar{\gamma}}\:e^{-m^P_{1/2} t}\nn
\label{eq:c2btlarge}
\eea
where the projection operators are defined by
\be
\begin{array}{ll}
P_-=\dfrac{1+\g_0}{2}\,,&\quad P_+=\dfrac{1-\g_0}{2},\\
P_{3/2}^{ij}=g^{ij}- \dfrac{1}{3}\g^i\,\g^j\,, &\quad
 P_{1/2}^{ij}=\dfrac{1}{3}\g^i\,\g^j\,.
\end{array}
\ee 
Contributions of negative parity states are removed by projection
with $P_+$. The negative parity states can, in principle, be
detected by using the projector $P_-$, but in our
simulation they are much noisier.
\begin{figure}
\begin{center}
\epsfxsize=\textwidth
\epsfbox{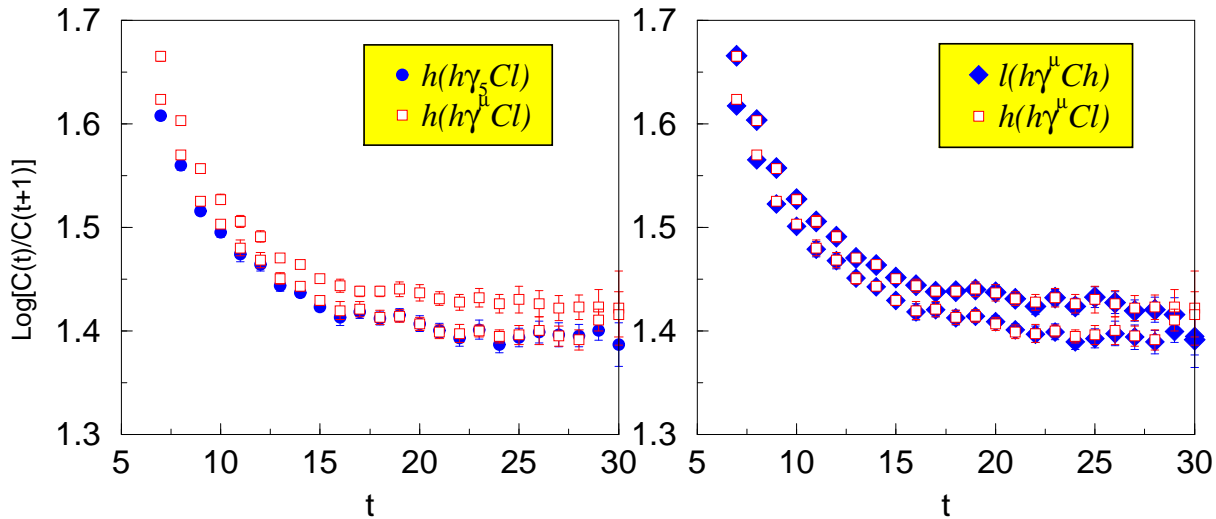}
\end{center}
\caption{Comparison of the effective mass plots for the double heavy
operators $J_\g$, $J^\mu_\g$ and $\widetilde J^\mu_\g$, with
$\kappa_h=0.1222$ and $\kappa_l=0.1351$. In each plot the upper points
are for spin-$3/2$ and the lower points for spin-$1/2$. The spin-$1/2$
plateaus are the the same for $J_\g$ and $J^\mu_\g$ (left) while both
plateaus coincide for the $J^\mu_\g$ and $\widetilde J^\mu_\g$
operators (right).}
\label{fig:hhl}
\end{figure}
We show an example of the signals from the operators $J_\g$,
$J^\mu_\g$ and $\widetilde J^\mu_\g$ in fig.~\ref{fig:hhl}. As
stressed above, spin-$1/2$ masses extracted using the three operators
are equal, while the choice between $J^\mu_\g$ and $\widetilde
J^\mu_\g$ makes no difference for the spin-$3/2$ mass.

For a baryon containing a single heavy quark, a common choice of operators is
\bea
{\cal O}_\gamma&=& \epsilon_{abc}\,h_{\gamma}^{a}\,\left({l^{b}_1}^T\,\gamma_5{\cal C} \,l^{c}_2\right),\quad s_{ll}=0,\\
{\cal O}^\mu_\gamma&=& \epsilon_{abc}\,h_{\gamma}^{a}\,\left({l^{b}_1}^T\,\gamma^\mu{\cal C} \,l^{c}_2\right),\quad s_{ll}=1,
\eea
for the states $s_{ll}=0$ and $s_{ll}=1$ in tab.~\ref{tab:Expc},
respectively.  In our simulation, the light quarks $l_1,\,l_2$ carry
different flavours but the same masses.

It should be noted that for baryons three different quarks, i.e., $h
l_1l_2$ (or $l h_1h_2$), these two operators correspond to different
physical spin-$1/2$ states with $s_{ll}=0$ and $1$ respectively, the
latter one often being denoted by a prime.  This is evident from the
experimental masses of $\Xi_c$ and $\Xi_c'$ in tab.~\ref{tab:Expc}.

\section{Details of the Simulation}
\label{sec:details}

Our simulation was made using the code FermiQCD~\cite{DiP01} on a PC
cluster. In this study $100$ quenched gauge configurations were
generated at $\beta = 6.2$ on a volume of $24^3\times 64$ with $1000$
heatbath steps for the thermalisation followed by $200$ heatbath steps
to separate each gauge configuration. These numbers were decided upon
after an autocorrelation study on the average plaquette values.

Four light quark propagators around the strange quark mass and three heavy
quark propagators around the charm were calculated using the following values
of the hopping parameters:
\begin{itemize}
\item $\kappa_l=0.1344\,,\, 0.1346\,,\, 0.1351\,,\, 0.1353$;
\item $\kappa_h=0.1240\,,\, 0.1233\,,\,0.1222$.
\end{itemize}
The propagators were generated by the Bistabilised Conjugate Gradient
method~\cite{Fro94} for the non-perturbatively improved clover
action~\cite{Lus96}.

Since the signal is satisfactory with local interpolating operators,
no smearing was required. The statistical errors were estimated by a
jacknife procedure, removing $10$ configurations at a time from the
ensemble.

\subsection{Lattice spacing and quark masses}

\begin{table}
\begin{center}
\begin{tabular}{|c|c|c|}
\hline 
$\kappa_l$ & $a m_P$ & $a m_V$ \\
\hline 
0.1344   & 0.300(2) & 0.397(4)  \\  
0.1346   & 0.276(2) & 0.383(5)  \\
0.1351   & 0.210(3) & 0.352(11) \\
0.1353   & 0.177(2) & 0.340(15) \\
\hline       
\end{tabular}
\end{center}
\caption{Light pseudoscalar and vector meson masses.
The fit interval is $[12-28]$. Our time 
counting starts from $0$.}
\label{tab:mlight}
\end{table}
\begin{figure}
\begin{center}
\epsfxsize=0.65\textwidth
\epsfbox{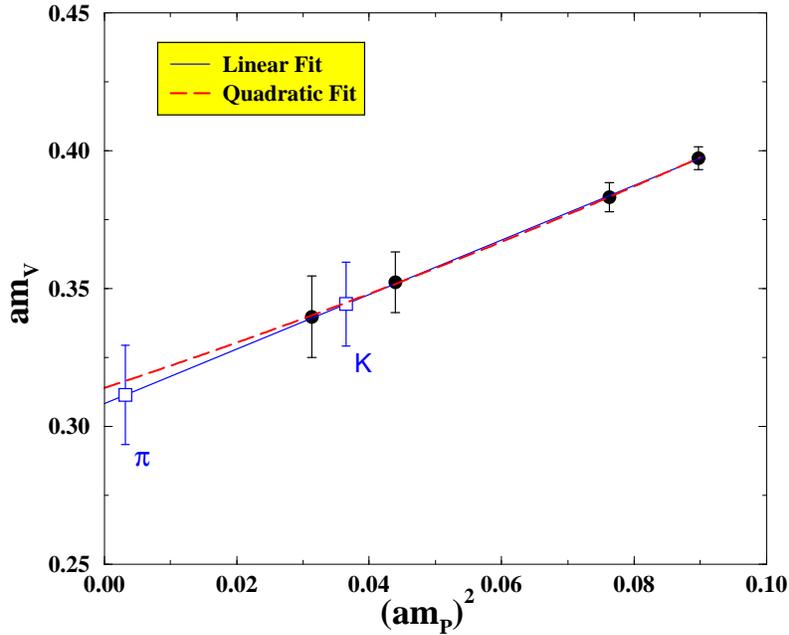}
\end{center}
\caption{Light vector masses as a function of squared light pseudoscalar masses. 
The interpolated kaon and extrapolated pion masses are also shown.}
\label{fig:lpm}
\end{figure}
To fix the lattice spacing, we used the {\em method of lattice planes}
\cite{All96}.  In other words, we perform the following fit to the light vector
and pseudoscalar masses in table~\ref{tab:mlight},
\begin{equation} 
a m_V= C + L \,(a m_P)^2.
\label{eq:lpm}
\end{equation}
This is shown in fig.~\ref{fig:lpm}. From the physical values of
$m_{K^*}$ and $m_K$, the inverse lattice spacing is found to be
\begin{equation}
a^{-1}=2.6(1)\, \gev.
\end{equation}
Terms of ${\cal O} ((a m_P)^4)$ in eq.~(\ref{eq:lpm}) turn out to be irrelevant
and do not affect the above estimate (compare the linear and quadratic fits in
fig.~\ref{fig:lpm}).
For illustration, the values of the pseudoscalar masses in
tab.~\ref{tab:mlight} converted to physical units are
\begin{equation}
m_P=\{779\,,\,716\,,\,546\,,\,459\}\,\mev.
\end{equation}
These span the kaon mass while the pion is instead quite far away. For this
reason, we interpolate for the strange quark and extrapolate for the up/down
masses.  This is also the reason for using $K,\,K^*$ to fix the lattice
spacing.

For the heavy sector, the $D_s$-meson mass is within our range of
simulation. This is evident once the heavy-light pseudoscalar masses in
tab.~\ref{tab:mheavy} are interpolated to the strange mass (through the lattice
plane method) and expressed in physical units
\begin{equation}
\label{eq:mhsphys}
m_{hs}=\{1.83\,,\,1.89\,,\,1.98\}\,\gev.
\end{equation}

\subsection{Analysis of the baryon masses}

\begin{figure}
\begin{center}
\epsfxsize=0.65\textwidth
\epsfbox{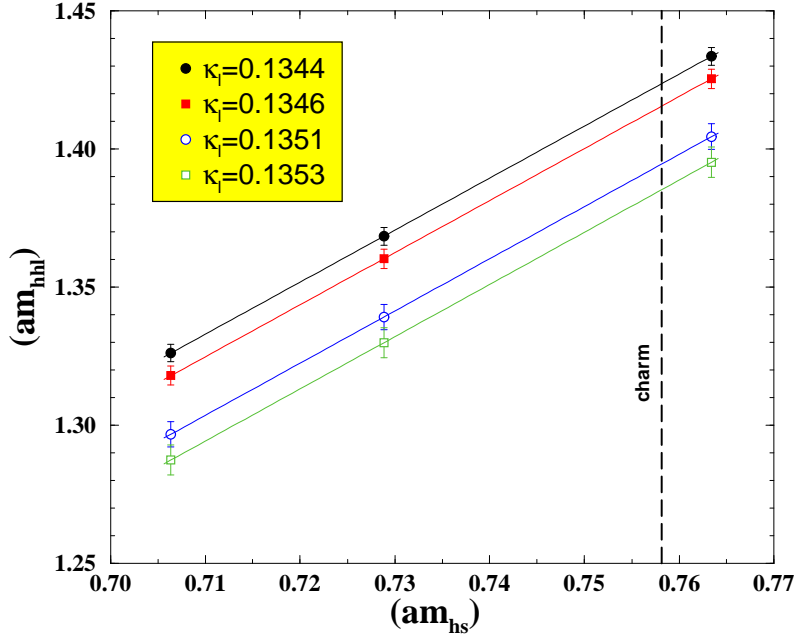}
\end{center}
\caption{Spin-$1/2$ double-heavy baryon masses for all $\kappa$
  combinations. For each $\kappa_l$ we fit the heavy quark mass
  dependence using the heavy-strange pseudoscalar meson mass. The fit
  function is given in equation~(\ref{eq:hhlfit}). The vertical dashed
  line indicates the $D_s$ meson mass (in lattice units) used to fix
  the masses of the $ccl$ spin-$1/2$ baryons.}
\label{fig:charm_xtra}
\end{figure}

\begin{figure}
\begin{center}
\epsfxsize=0.65\textwidth
\epsfbox{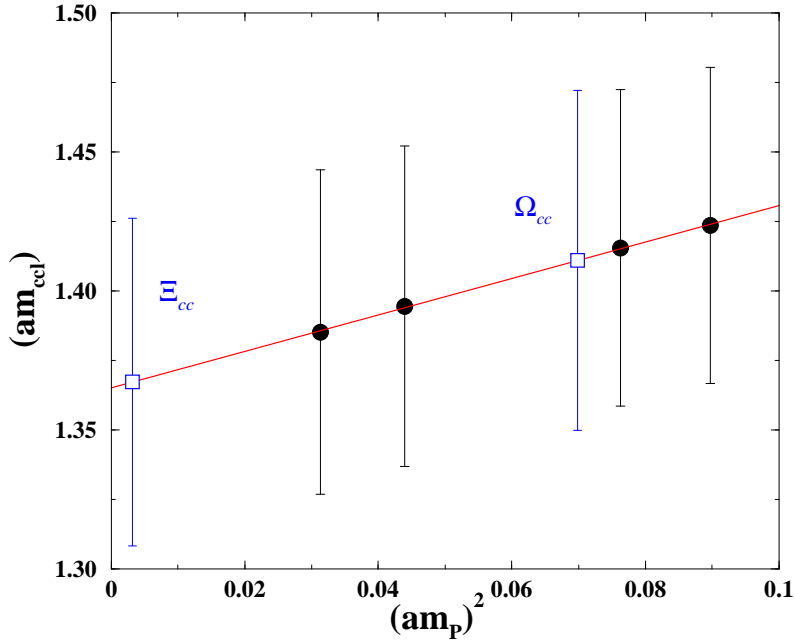}
\end{center}
\caption{Spin-$1/2$ double charm state masses as a function of the
square of the light pseudoscalar masses. The values at strange and
up/down masses are shown.}
\label{fig:uds_xtra}
\end{figure}

Since $\kappa_\mathrm{charm}$ is rather close to our third
$\kappa_h=0.1222$\footnote{A naive linear fit in $1/\kappa_h$ to the
masses in eq.~(\ref{eq:mhsphys}) gives
$\kappa_\mathrm{charm}=0.1224(9)$.}, as the first step in our
analysis, we interpolate the quantities of interest, {\em viz.} the
single and double heavy baryon masses, to the charm mass. In practice,
this procedure is implemented by doing for each $\kappa_l$ the
following fits
\begin{equation}
\label{eq:hhlfit}
a m_{hhl}= C_l + L_l\, a m_{hs}\,,\quad
a m_{hll}= C^\prime_l + L^\prime_l\, a m_{hs}.
\end{equation}
Quantities at the charm mass, $m_{ccl}$ and $m_{cll}$ are obtained by
putting  $m_{hs}=m_{D_s}$.   This interpolation is shown 
for the double heavy case in fig.~\ref{fig:charm_xtra}.
\begin{table}
\begin{center}
\def\arraystretch{1.35}
\begin{tabular}{|c|c|c|c|c|c|c|}
\hline 
&  & \multicolumn{3}{c|}{$a m_{hll'}$}  & \multicolumn{2}{c|}{$a m_{hhl}$} \\ \cline{3-4}\cline{5-7}
$\kappa_h-\kappa_l$ & $a m_P$ & $J^P=\frac{1}{2}^{+}$ & $J^P=\frac{1}{2}^{+}$ & $J^P=\frac{3}{2}^{+}$ & $J^P=\frac{1}{2}^{+}$ & $J^P=\frac{3}{2}^{+}$ \\
& & $s_{ll'}=0$ & $s_{ll'}=1$ & $s_{ll'}=1$ & $s_{hh}=1$ & $s_{hh}=1$ \\
\hline 
0.1240-0.1344   & 0.718(2)  & 0.954(5) & 0.988(6) & 1.008(6) & 1.326(3) & 1.354(3) \\  
0.1233-0.1344   & 0.740(2)  & 0.975(5) & 1.010(6) & 1.029(6) & 1.368(3) & 1.395(3) \\  
0.1222-0.1344   & 0.775(2)  & 1.008(6) & 1.044(6) & 1.062(6) & 1.433(3) & 1.459(3) \\  
$\kappa_\mathrm{charm}$-0.1344   &   &1.003(28) & 1.039(33) & 1.057(32) &  1.055(31) & 1.442(57) \\  
\hline
0.1240-0.1346   & 0.710(2)  & 0.934(6) & 0.972(7) & 0.992(7) &  1.318(4) & 1.347(4) \\  
0.1233-0.1346   & 0.733(2)  & 0.956(6) & 0.994(7) & 1.013(7) &  1.360(4) & 1.388(3) \\  
0.1222-0.1346   & 0.767(2)  & 0.989(6) & 1.028(7) & 1.046(7) &  1.425(3) & 1.452(3) \\  
$\kappa_\mathrm{charm}$-0.1346   &  & 0.984(28)& 1.023(34) & 1.041(33) &  1.416(57) & 1.442(57) \\  
\hline
0.1240-0.1351   & 0.691(3) & 0.878(10) & 0.929(13) & 0.945(10) &  1.297(4) & 1.329(5) \\  
0.1233-0.1351   & 0.714(3) & 0.900(10) & 0.951(13) & 0.966(10) &  1.339(5) & 1.370(5) \\  
0.1222-0.1351   & 0.748(3) & 0.934(10) & 0.984(14) & 0.998(10) &  1.404(5) & 1.434(5) \\  
$\kappa_\mathrm{charm}$-0.1351   &  &0.928(27) & 0.979(39) & 0.993(34) & 1.395(58) & 1.425(58) \\  
\hline
0.1240-0.1353   & 0.683(3) & 0.854(13) & 0.903(17) & 0.915(12) &  1.287(5) & 1.322(6) \\  
0.1233-0.1353   & 0.706(3) & 0.876(13) & 0.923(17) & 0.935(12) &  1.330(5) & 1.363(6) \\  
0.1222-0.1353   & 0.740(3) & 0.910(13) & 0.956(18) & 0.967(12) &  1.395(6) & 1.427(6) \\  
$\kappa_\mathrm{charm}$-0.1353   &  &0.904(28) & 0.951(43) & 0.962(34) & 1.385(58) & 1.418(58) \\  
\hline
\end{tabular}
\end{center}
\caption{Double and single-heavy baryon masses in lattice units, together with
pseudoscalar masses.  The fit intervals are $[16-28]$ for double and $[15-25]$
for single-heavy baryons.}
\label{tab:mheavy}
\end{table}
With the charm mass fixed, the light quark mass dependence is studied using
\begin{equation}
a m_{cll}= A + B\, (a m_P)^2\,,\quad
a m_{ccl}= A' + B'\, (a m_P)^2.
\end{equation}
This fit is shown for the spin-$1/2$ double charm case in
fig.~\ref{fig:uds_xtra}. The masses of charmed baryons containing
strange and/or up/down quarks are obtained by the following
substitutions for $m_P$ in the above equations:
\begin{itemize}
\item $m_P=m_\pi$ for $m_{cud}$, $m_{ccu}$;
\item $m_P=m_{K}$ for $m_{csu}$; 
\item $m_P=m_{\eta_{ss}}$ for $m_{css}$, $m_{ccs}$.
\end{itemize}
where $m_{\eta_{ss}}^2 = 2m_K^2 -m_\pi^2$. In the second case, we
suppose that $SU(3)$ breaking terms are negligible and obtain
our estimate from states containing two mass-degenerate light
quarks~\cite{All96,Ali00}.

\section{Results and Discussion}
\label{sec:results}

Here we recall our final values for the double-charm baryon masses.
\be
\begin{array}{ll}
\Xi_{cc}=\xii\,\mev &\quad \Omega_{cc}=\om\,\mev \\
\Xi^*_{cc}=\xiis\,\mev & \quad\Omega^*_{cc}=\oms\,\mev
\end{array}
\ee
The first error is statistical. The second error is systematic,
estimated by combining in quadrature the effects of the following
variations in our analysis:
\begin{itemize}
\item changing the time fit-ranges --- this contributes up to $35\%$ of
  the quoted error;
\item using single or double exponential fits --- we saw no change in our
  lowest state masses;
\item linear versus quadratic chiral extrapolations --- in the worst case
  this gives three-quarters of the quoted error;
\item interchanging the order of light quark
  extrapolations and charm quark interpolation --- this produces no
  change in our results.
\end{itemize}
Only one volume and lattice spacing was studied; investigation of
discretization errors, the continuum limit and finite volume effects
are not addressed. To account for these (and the effects of
quenching), the third quoted error is found by rescaling our masses
using the experimental $\Lambda_c$ mass.

The $\Xi_{cc}$ mass is in good agreement with the experimental
value~\cite{Mattson02}
\be
(\Xi_{cc})_\mathrm{expt}=3519\pm 1\pm 5\,\mev
\ee
Other masses are consistent with the lattice estimates using
NRQCD~\cite{Mat02} or D234~\cite{Lew01} actions. For recent
estimates in quark models or QCD Sum Rules we refer the reader
to~\cite{Nar02} and~\cite{KK01} respectively.
\begin{table}
\begin{center}
\def\arraystretch{1.3}
\begin{tabular}{|ccc|}
\hline
 & This work [MeV] & Expt [MeV] \\
\hline
$\Lambda_{c}$ & $2227(50)(57)(58)$  &  $2285(1)$\\
$\Xi_{c}$     & $2374(34)(23)(61)$  &  $2469(1)$\\
\hline
$\Sigma_{c}$  	   & $2377(38)(84)(62)$  & $2452(1)$\\
$\Xi^{\prime}_{c}$ & $2502(26)(40)(65)$  & $2575(3)$\\
$\Omega_{c}$       &  $2627(16)(48)(68)$ & $2698(3)$\\
\hline
$\Sigma^{*}_{c}$ &  $2396(42)(122)(62)$  & $2518(2)$ \\
$\Xi^{*}_{c}$    &  $2532(31)(62)(66)$  & $2646(2)$\\
$\Omega^{*}_{c}$ &  $2669(21)(26)(70)$  & \\
\hline
\end{tabular}
\end{center}
\caption{Our estimates for the single charm baryon masses compared to
experimental values.}
\label{tab:mcllphys}
\end{table}
For completeness, our estimates for the single charm baryon masses are
given in tab.~\ref{tab:mcllphys} along with the experimental
results. Values turn out to be compatible with previous lattice
calculations~\cite{Bow96,Mat02,Ali00}. It may be noted that in
ref.~\cite{Bow96}, a perturbative value for the coefficient
$c_{SW}$ was used in the clover action.

We now turn to the baryon and meson spin-splittings. Our results for
these are given in tab.~\ref{tab:dmphys}. The values are obtained
either from the difference in individually fitted masses (labelled
``Diff'' in the table), or by directly fitting a ratio of correlators
(labelled ``Ratio'' in the table). When using the ratio the noise
starts to dominate earlier so we restrict our fit to a shorter
time-slice window. For the baryons we find a better signal using the
ratio method and the difference between the two approaches becomes
more apparent as we move away from our region of simulation to lighter
quarks. We use the numbers from the ratio as our best estimates.

For the double-charm baryons we observe a good signal for non-zero
splittings. For the single-charm baryons, where experimental data is
available, our results are compatible. This distinguishes our results
from earlier ones using a less-improved clover action~\cite{Bow96}.
Our values are also compatible with those found using the
D234~\cite{Lew01} or NRQCD~\cite{Mat02} actions. For the mesons too
our results are compatible with experiment: this improved agreement is
also found in other recent non-perturbatively improved clover
simulations~\cite{Bec98,Bow00}.

\begin{table}
\begin{center}
\def\arraystretch{1.3}
\begin{tabular}{|cccc|}
\hline
 &  Diff [MeV] & Ratio [MeV] & Expt [MeV] \\
\hline
$\Xi^{*}_{cc}-\Xi_{cc}$          &$\dxiccd$ & $\dxiccr$ & \\
$\Omega^{*}_{cc}-\Omega_{cc}$    &$\domccd$ & $\domccr$ & \\
\hline
$\Sigma^{*}_{c}-\Sigma_{c}$      &$\dsigcd$ & $\dsigcr$ & 66(2) \\ 
$\Xi^{*}_{c}-\Xi^{\prime}_{c}$   &$\dxicd$  & $\dxicr$  & 71(3) \\
$\Omega^{*}_{c}-\Omega_{c}$      &$\domcd$  & $\domcr$  & \\
\hline
$D^* - D$                        &          & 127(14)(1)(3)  & 142(2) \\
$D_s^* - D_s$                    &          & 123(11)(1)(3)  & 138(2) \\
\hline
\end{tabular}
\end{center}
\caption{Our results for the single- and double-charm mass
  splittings.}
\label{tab:dmphys} 
\end{table}

The predictions are more precise for $(\Omega^{*}_{c},\Omega_{c})$ and
the double charm spin doublets $(\Omega^{*}_{cc},\Omega_{cc})$ and
$(\Xi^{*}_{cc},\Xi_{cc})$, where less extrapolation is needed, but
experimental numbers are still awaited.

\section{Conclusion}
\label{sec:conclusion}

We have presented exploratory quenched lattice results for double
charm baryon masses. These have drawn attention after the experimental
observation of the first double charm state last year. The calculation
is done with non-perturbatively ${\cal O}(a)$-improved Wilson fermions
at $\beta=6.2$ and on a large lattice. Good signals for the ground
states are observed without recourse to smearing. In addition, we have
reported the masses of single charm baryons. The calculated masses
look quite reasonable. We see a definite signal for non-zero baryon
and meson spin splittings. To improve our lattice calculations, a
finer lattice spacing is necessary together with an examination of
chiral logarithms in the light extrapolations and a simulation with
dynamical quarks.

Experimental observations of the remaining double charmed baryon
states and, in particular their spin-splittings, would allow the
lattice predictions to be checked.

\section*{Acknowledgments}
We would like to thank Massimo Di Pierro for his help in using the
FermiQCD code. We also thank the Iridis parallel computing team at
University of Southampton, in particular, Ivan Wolton, Ian Hardy and
Oz Parchment for their computing support. We thank Craig McNeile and
Chris Maynard for useful discussions and cross-checks; Damir
Becirevic, Andreas Kronfeld, Vittorio Lubicz and Cecilia Tarantino for
comments. The work of ASBT is supported by a Commonwealth Scholarship.
Work partially supported by the European Community's Human Potential
Programme under HPRN-CT-2000-00145 Hadrons/Lattice QCD.


\end{document}